\documentclass[aps,prb,twocolumn,superscriptaddress]{revtex4}
\makeatletter

\newcommand{\Rmnum}[1]{\expandafter\@slowromancap\romannumeral #1@}
\makeatother

\usepackage{eurosym}
\usepackage{amsfonts}
\usepackage{amssymb}
\usepackage{amsmath}
\usepackage{graphicx}
\usepackage{color}
\begin{document}

\title{Enhancing the thermoelectric performance of a HfS$_{2}$ monolayer through valley engineering}

\author{H. Y. Lv}

\affiliation{Key Laboratory of Materials Physics, Institute of Solid State Physics, Chinese Academy of Sciences, Hefei 230031, People's Republic of China}

\author{W. J. Lu}
\email[Corresponding author: ]{wjlu@issp.ac.cn}
\affiliation{Key Laboratory of Materials Physics, Institute of Solid State Physics, Chinese Academy of Sciences, Hefei 230031, People's Republic of China}

\author{X. Luo}

\affiliation{Key Laboratory of Materials Physics, Institute of Solid State Physics, Chinese Academy of Sciences, Hefei 230031, People's Republic of China}

\author{H. Y. Lu}

\affiliation{School of Physics and Electronic Information, Huaibei Normal University, Huaibei 235000, People's Republic of China}

\author{X. B. Zhu}

\affiliation{Key Laboratory of Materials Physics, Institute of Solid State Physics, Chinese Academy of Sciences, Hefei 230031, People's Republic of China}

\author{Y. P. Sun}
\email[Corresponding author: ]{ypsun@issp.ac.cn}
\affiliation{Key Laboratory of Materials Physics, Institute of Solid State Physics, Chinese Academy of Sciences, Hefei 230031, People's Republic of China}
\affiliation{High Magnetic Field Laboratory, Chinese Academy of Sciences, Hefei 230031, People's Republic of China}
\affiliation{Collaborative Innovation Center of Advanced Microstructures, Nanjing University, Nanjing, 210093, People's Republic of China}

\makeatletter

%%%%%%%%%%%%%%%%%%%%%%%%%%%%%% LyX specific LaTeX commands.

\begin{abstract}
The electronic, phonon, and thermoelectric properties of a two-dimensional HfS$_{2}$ monolayer are investigated by using the first-principles calculations combined with the Boltzmann transport theory. The band valleys of the HfS$_{2}$ monolayer can be effectively tuned by the applied biaxial strain. The Seebeck coefficient and therefore the peak value of the power factor (with the relaxation time inserted) increase when the degeneracy of the band valleys is increased by the strain. When no strain is applied, the HfS$_{2}$ monolayer is an excellent n-type thermoelectric material, while the thermoelectric performance of the p-type doped one is poor. The applied tensile strain of 6\% can increase the room-temperature $ZT$ value of the p-type doped system to 3.67, which is five times larger than that of the unstrained one. The much more balanced $ZT$ values of the p- and n-type doping are favorable for fabrication of both p- and n-legs of thermoelectric modules. Our results indicate that the thermoelectric performance of the HfS$_{2}$ monolayer can be greatly improved by the valley engineering through the method of strain.

\end{abstract}
\maketitle

\section{INTRODUCTION}
With most of our used energy being lost as waste heat, there is a growing need for high-performance thermoelectric materials that can directly and reversibly convert heat into electricity. The efficiency of a thermoelectric material is determined by the dimensionless figure of merit $ZT=S^2\sigma T/(\kappa_e+\kappa_p)$, where $S$ is the Seebeck coefficient, $\sigma$ is the electrical conductivity, $T$ is the temperature, and $\kappa_e$ and $\kappa_p$ are the electronic and phonon thermal conductivities, respectively. To obtain a high $ZT$ value, one must try to increase the power factor (PF=$S^2\sigma$) and/or decrease the thermal conductivity. However, the transport coefficients ($S$, $\sigma$, $\kappa_e$ and $\kappa_p$) are not independently tunable because they all depend strongly on the details of the band structure and the scattering of the charge carriers. As a result, it is still a challenge to obtain high $ZT$ thermoelectric materials, hindering their wide applications.

Several successful concepts have been developed to increase the $ZT$ value, such as the use of low dimensionality,\cite{MS1,MS2} which could enhance the PF due to the sharper density of states (DOS) near the Fermi energy or reduce the thermal conductivity originating from the increased phonon scattering. Therefore, many efforts have been devoted to search for low-dimensional thermoelectric materials. Due to the weak van der Waals interactions between the neighboring layers, the layered materials serve as ideal candidates which can be readily exfoliated to two-dimensional (2D) films.\cite{synthesis1,synthesis2} On the other hand, a high valley degeneracy in the electronic structure also contributes to the enhancement of the thermoelectric performance.\cite{valley1,valley2,valley3} Such degeneracy can be engineered by tuning the doping and composition in a bulk material.\cite{valley-bulk1,valley-bulk2,valley-bulk3} If we could combine the above two concepts together, namely, introduce the high valley degeneracy into a low dimensional thermoelectric material, a large $ZT$ value may be achieved.

\begin{figure}[htpb]
\centering
\includegraphics[width=1.0\columnwidth]{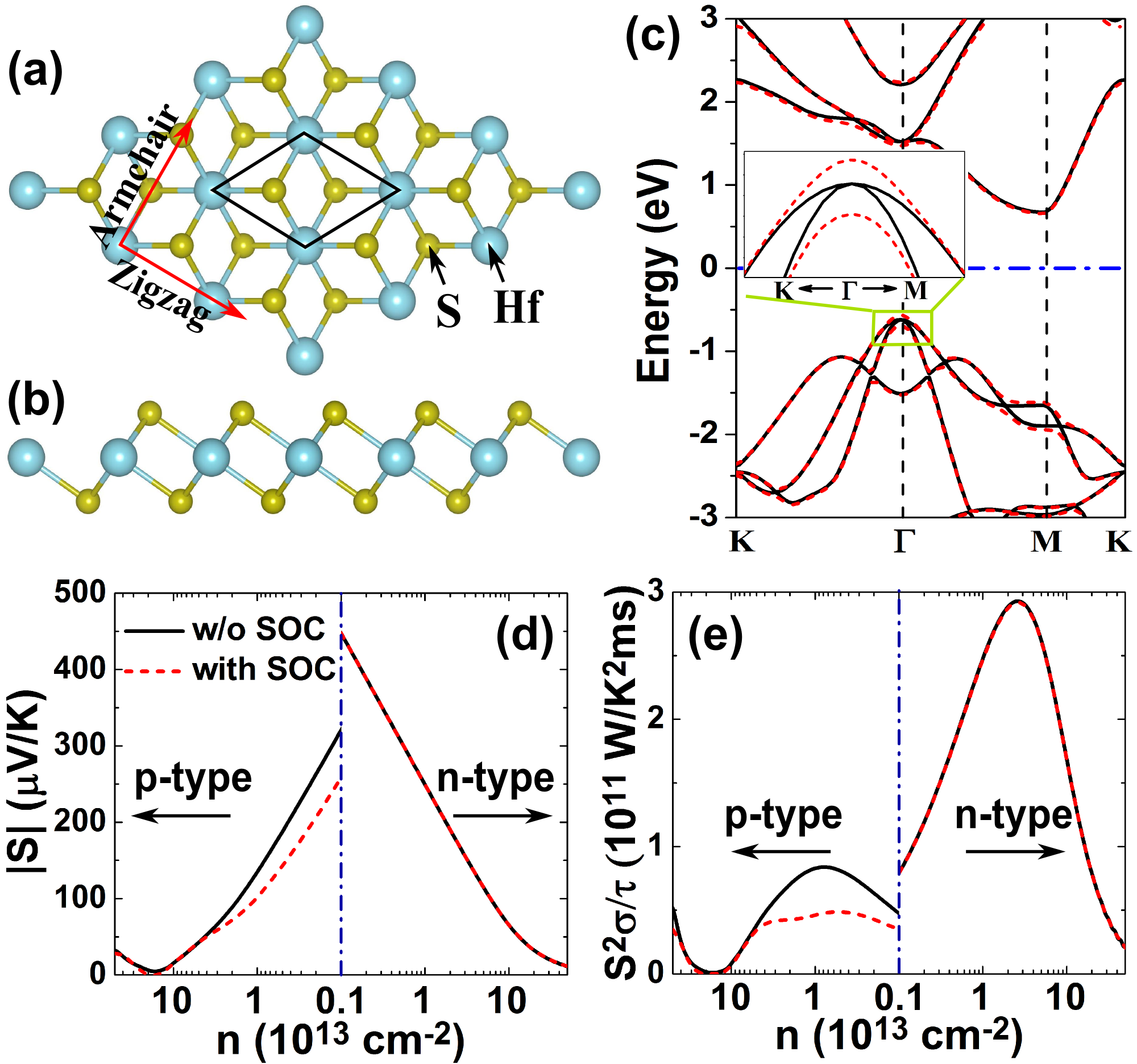}\caption{\label{fig1-structure and energy band}(a) Top and (b) side views of the HfS$_{2}$ monolayer. The black line in (a) denotes the primitive cell used in our calculations. (c) Band structures of the HfS$_{2}$ monolayer without (solid black line) and with (dashed red line) spin-orbit coupling (SOC). The inset is a magnified view of the region marked by the rectangular outline. (d) Absolute value of the Seebeck coefficient and (e) power factor (PF) as a function of the carrier concentration. The solid black and dashed red lines represent the results without and with SOC, respectively.}
\end{figure}

\begin{figure}[htpb]
\centering
\includegraphics[width=1.0\columnwidth]{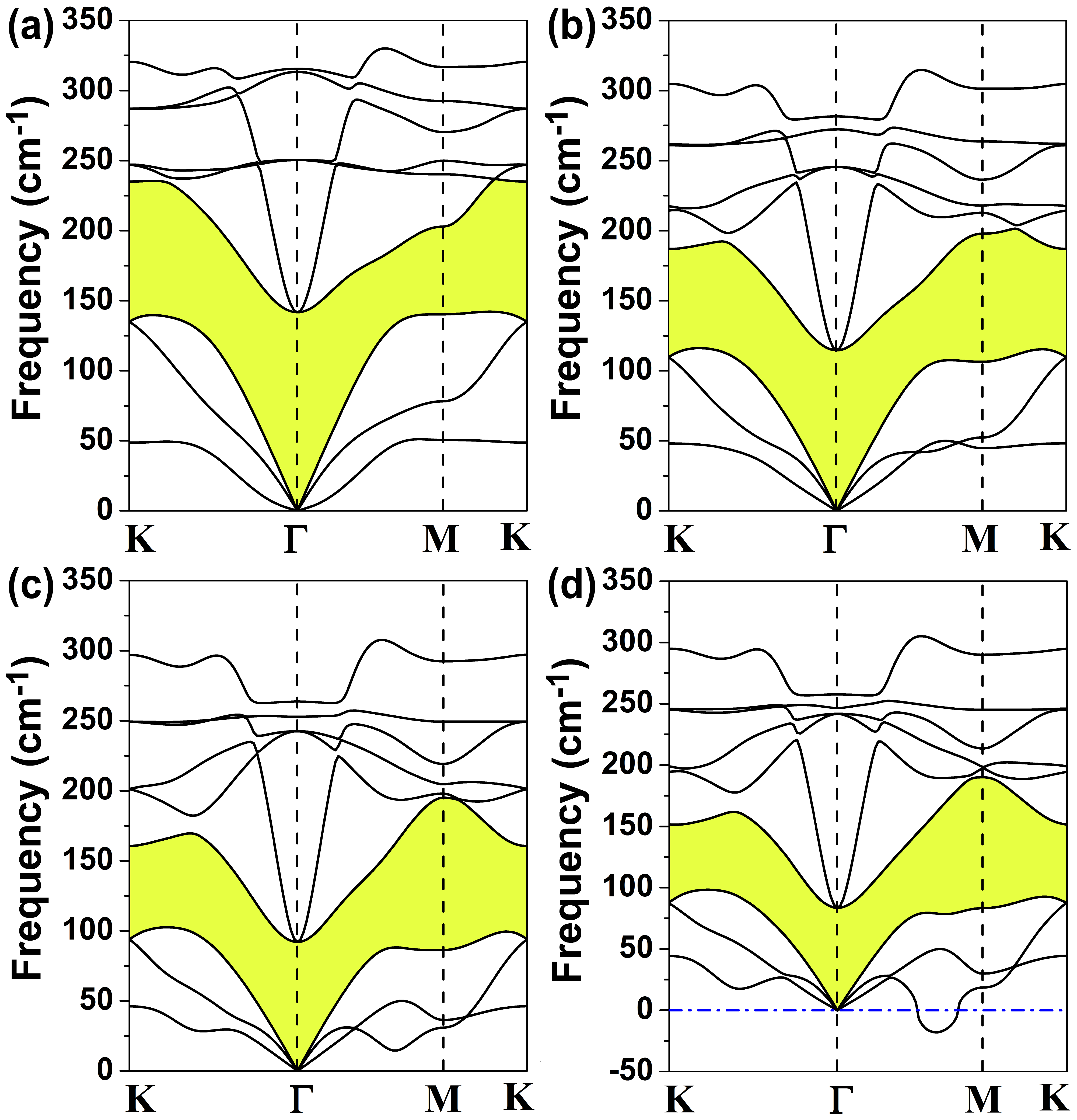}\caption{\label{fig2-band struct}Phonon spectra of the HfS$_{2}$ monolayer under different external strains of (a) 0\%, (b) 6\%, (c) 9\%, and (d) 10\%. The gap between acoustical and optical phonon bands is highlighted in yellow color.}
\end{figure}

Transition metal dichalcogenides (TMDCs), which have the formula M$X_2$ (M = transition metal; $X$ = S, Se, or Te), represent a large family of layered materials. TMDCs are presently being intensively researched due to their diverse and attractive properties.\cite{TMDC1,TMDC2,TMDC3} Some of them, TiS$_2$ for example,\cite{TiS2-bulk} were found to be good thermoelectric materials. When exfoliated into film, the 2D TiS$_2$ was predicted to have much improved thermoelectric performance compared with its bulk counterpart when the thickness fell below 14 layers.\cite{TiS2-layer} HfS$_2$ is another typical TMDC. It was theoretically predicted that the lattice thermal conductivity of the bulk HfS$_2$ was about 9 Wm$^{-1}$K$^{-1}$ at 300 K,\cite{HfS2-bulk} which is much smaller than those of the MoS$_2$\cite{MoS2-bulk-kappa} and WS$_2$,\cite{WS2-bulk-kappa} making the HfS$_2$ system a promising thermoelectric material. However, the room-temperature $ZT$ value of the bulk HfS$_2$ was about 0.06,\cite{HfS2-bulk}, still far away from the requirement of the thermoelectric application.\cite{ZT-require} If using the method of low dimensionality, the thermoelectric performance of the HfS$_2$ system may be further enhanced. In this work, we investigate the thermoelectric properties of the 2D HfS$_2$ monolayer. It is demonstrated that the HfS$_2$ monolayer is an excellent n-type thermoelectric material, with much improved thermoelectric performance compared with the bulk. Furthermore, the band valleys in the HfS$_2$ monolayer can be effectively tuned by the external strain, and the Seebeck coefficient and therefore the peak value of the PF are greatly increased. At the strain of 6\%, where the degeneracy of the valence band valleys reach the maximum, the $ZT$ value of the p-type doped system is dramatically increased.

\section{Computational details}

Our calculations were performed within the framework of the density functional theory (DFT), as implemented in the ABINIT code.\cite{ABINIT1,ABINIT2,ABINIT3} The exchange correlation energy was in the form of Perdew-Burke-Ernzerhof (PBE)\cite{PBE} with generalized gradient approximation (GGA). The Brillouin zone was sampled with a $12\times12\times1$ Monkhorst-Pack $k$ mesh. The cutoff energy for the plane-wave expansion was set to be 600 eV. Based on the electronic structure, the electronic transport coefficients are derived by using the semiclassical Boltzmann theory\cite{Boltzmann} within the relaxation-time approximation, and doping is treated by the rigid-band model.\cite{rigid-band} The electronic thermal conductivity $\kappa_e$ is calculated using the Wiedemann-Franz law $\kappa_e =L\sigma T$, where $L$ is the Lorenz number. In this work, we use a Lorenz number of $1.50\times10^{-8}$ W/K$^2$.\cite{Lorenz}

In the calculation of the phonon dispersion, the density functional perturbation theory (DFPT) as implemented in the VASP package\cite{vasp1,vasp2,vasp3} was used to calculate the force constant matrices. A $5\times5\times1$ supercell was used and the phonon frequencies were obtained by the PHONOPY code.\cite{phonopy} The lattice thermal conductivity was calculated by solving the phonon Boltzmann transport equation within the relaxation time approximation, as implemented in the ShengBTE code.\cite{ShengBTE} The second order harmonic and third order anharmonic interatomic force constants (IFCs) were calculated by using $5\times5\times1$ supercell with $2\times2\times1$ Monkhorst-Pack $k$ meshes and $4\times4\times1$ supercell with $\Gamma$ point, respectively.  The interactions up to third-nearest neighbors were considered when calculating the third order IFCs.

\begin{figure*}[ht]
\centering
\includegraphics[width=1.5\columnwidth]{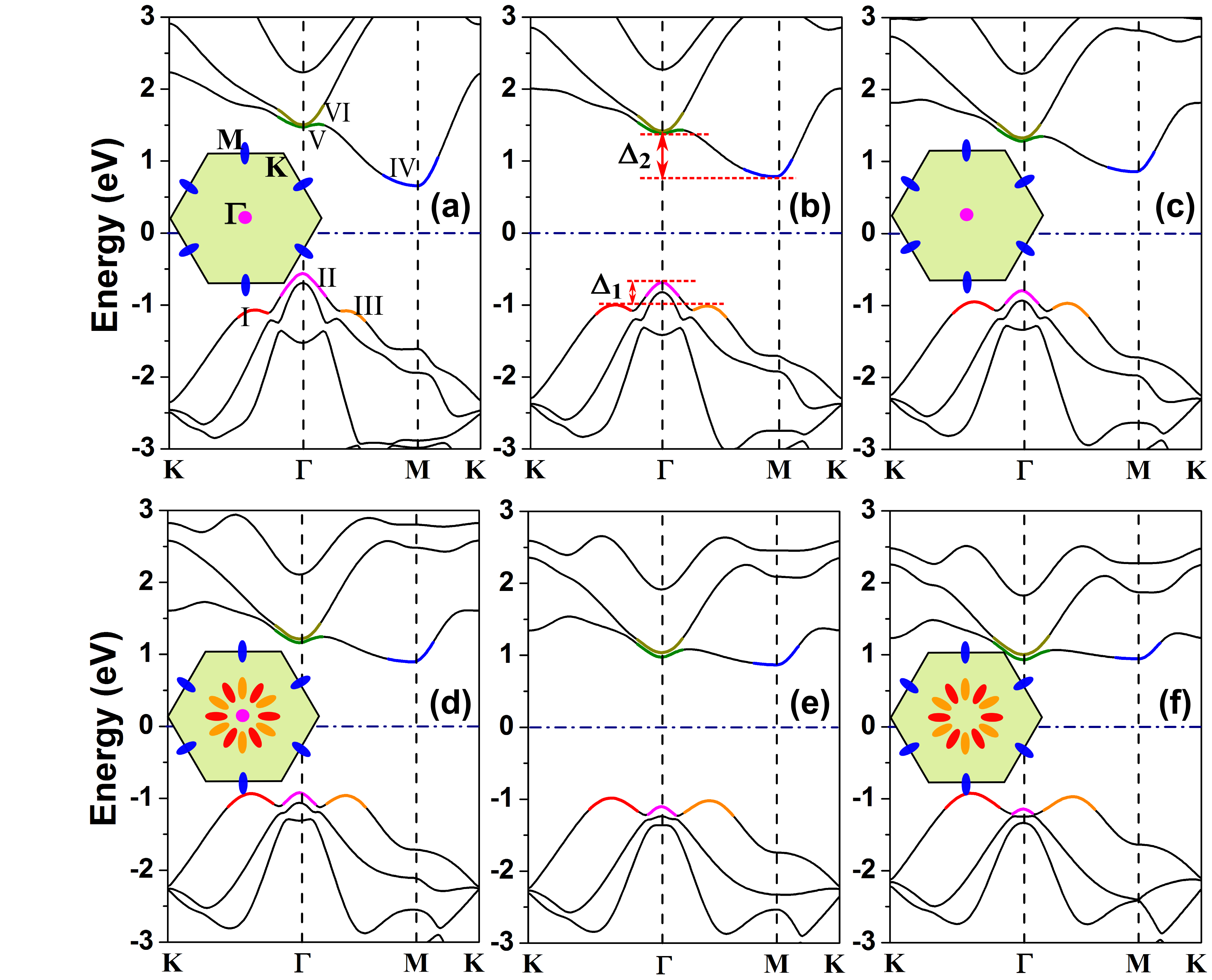}\caption{\label{fig2-band struct}Band structures of the HfS$_{2}$ monolayer under different external strains of (a) 0\%, (b) 2\%, (c) 4\%, (d) 6\%, (e) 8\%, and (f) 9\%. The three valence band valleys are highlighted in different color, denoted by I, II, and III, respectively; the three conduction band valleys are denoted by IV, V, and VI, respectively. For the unstrained system, the VBM and CBM are determined by valleys II and IV located at $\Gamma$ and $M$ points, respectively, which are schematically drawn in the Brillouin zone in the inset of (a). The same are shown in the insets of (c), (d), and (f) for the systems under the strains of 4\%, 6\%, and 9\%, respectively. At the strain of 6\%, the three valence band valleys converge in energy and the degeneracy reaches the maximum, as shown in the inset of (d).}
\end{figure*}

\section{RESULTS AND DISCUSSION}

The bulk HfS$_{2}$ crystallizes in the 1$T$-CdI$_{2}$ structure with the space group of $P\bar{3}m$1.\cite{bulk-structure} Due to the weak van der Waals interaction between the layers, few-layer HfS$_{2}$ could be exfoliated from the bulk structure.\cite{few-layer HfS2} The top and side views of the HfS$_{2}$ monolayer are shown in Figs. 1(a) and (b), respectively. After full relaxation, the lattice parameters are calculated to be $a=b=3.639$ {\AA}, slightly larger than the in-plane parameters of the HfS$_{2}$ bulk.\cite{bulk-structure} The structural stability is investigated by calculating the phonon spectrum, which is shown in Fig. 2(a). There is no imaginary frequency in the phonon dispersion, indicating that the HfS$_{2}$ monolayer is stable. The result agrees well with the previous report,\cite{monolayer-phonon} confirming the reliability of our calculations. We next consider the electronic properties. The band structures without and with spin-orbit coupling (SOC) included are shown in Fig. 1(c), displayed by solid black and short dashed red lines, respectively. The HfS$_{2}$ monolayer is an indirect-band-gap semiconductor, with the valence band maximum (VBM) and conduction band minimum (CBM) located at the $\Gamma$ and $M$ points, respectively. The band splitting due to the SOC is observed. In particular, at $\Gamma$ point, the splitting energy $\Delta_{\mbox{so}}$ of the VBM is 0.13 eV, thus the degeneracy of the two valleys near the VBM changes from 2 without SOC to 1 when SOC is included. However, the degeneracy of the conduction band valley near the CBM is unchanged. The band gap is calculated to be 1.22 (1.29) eV with (without) SOC included. The electronic transport coefficients ($S$, $S^2\sigma/\tau$) are shown in Figs. 1(d) and (e), respectively. We can see that for the p-type doping, the Seebeck coefficient and therefore the PF (with relaxation time $\tau$ inserted) are decreased due to the SOC, while for the n-type doping, they keep unchanged. If we notice the change of the band structure due to the SOC, we can deduce that the decrease of the Seebeck coefficient may be caused by the decrease of the valley degeneracy. Therefore, a possible way to optimize the PF is to engineer the band valleys.

Strain is an effective way to tune the electronic structure of the 2D material. In the following, a biaxial tensile strain, which is defined as $\varepsilon=(a-a_0)/a_0\times100\%$, is applied to the HfS$_{2}$ monolayer. To see how large strain the HfS$_{2}$ monolayer can withstand, we calculate the phonon spectra when the strain is increased up to 10\%. The results for the strains of 6\%, 9\%, and 10\% are displayed in Figs. 2(b), (c) and (d), respectively. No imaginary frequencies are observed in the phonon spectra until the strain increases up to 9\%. When the strain reaches 10\%, small imaginary frequency appears along the $\Gamma-M$ direction, thus the system tends to be unstable. Therefore, the largest strain that the HfS$_{2}$ monolayer can withstand is 9\%. The HfS$_{2}$ monolayer can withstand a relatively large strain, mainly originating from the special sandwich structure. When the strain is applied, the outer S atom layers move inside, so the thickness of the HfS$_{2}$ monolayer is decreased, while the bond length between Hf and S atoms is slightly altered.

\begin{figure}[ht]
\centering
\includegraphics[width=0.8\columnwidth]{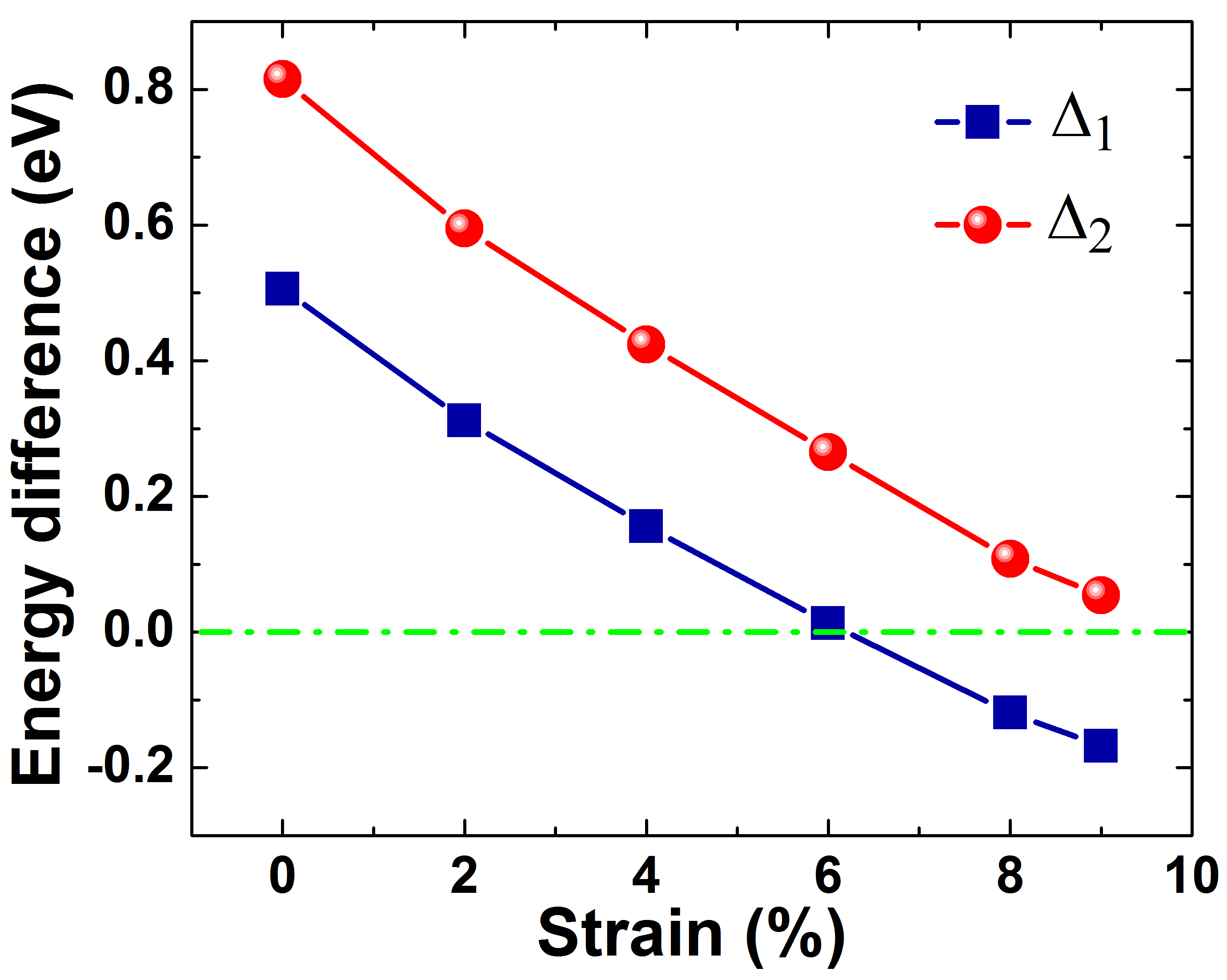}\caption{\label{fig3-energy difference}Energy difference between the three valence band valleys ($\Delta_1$) and between the three conduction band valleys ($\Delta_2$) as a function of the applied strain.}
\end{figure}

\begin{figure*}[htpb]
\centering
\includegraphics[width=1.4\columnwidth]{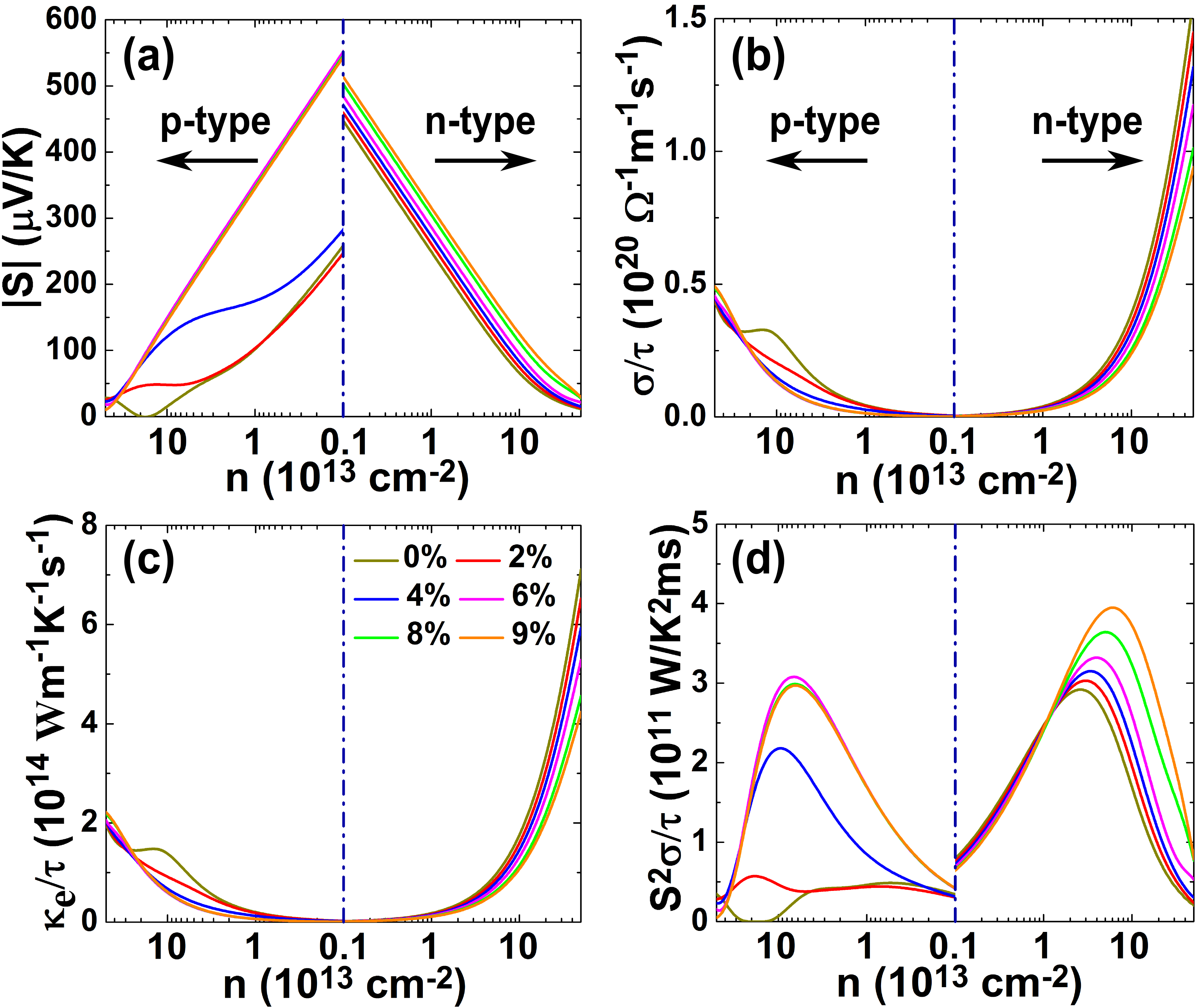}\caption{\label{fig5-band struct}Carrer concentration dependence of (a) absolute value of the Seebeck coefficient, (b) electrical conductivity, (c) electronic thermal conductivity, and (d) PF of the HfS$_{2}$ monolayer at 300 K under different external strains. The left and right panels are the p- and n-type doping, respectively.}
\end{figure*}

Next, the strain effect on the electronic properties is investigated. When the doping concentration is not very high, the electronic transport coefficients of the p- and n-type doped systems are determined by the bands near the VBM and CBM, respectively. For the valence bands, the three valleys highlighted in different color in Fig. 3(a) are denoted as I, II, and III, respectively, while the three conduction band valleys are denoted as IV, V, and VI. When no strain is applied, the energy of band valley II is the largest among the three valence band valleys and the energy difference between valleys II and I (III) is denoted by $\Delta_1$, as shown in Fig. 3(b). The energy of the valley IV is the smallest among the conduction band valleys and the energy difference between IV and V (VI) is denoted by $\Delta_2$. When the strain is applied, for the valence band valleys, the valleys I and III are elevated gradually when the strain is increased, while the valley II is gradually lowered, resulting in the decrease of $\Delta_1$, as displayed in Fig. 4. At the strain of 6\%, the three valleys converge in energy and $\Delta_1$ is decreased to zero. At this strain condition, the degeneracy of the valence band valley reaches the maximum, as displayed by the inset of Fig. 3(d). When further increasing the strain, the energy of valleys I and III becomes larger than that of the valley II, and the degeneracy is reduced, as demonstrated by the inset of Fig. 3(f). For the conduction band valleys, as the strain is increased, the valley IV is gradually elevated, while the valleys V and VI are lowered, therefore, the $\Delta_2$ decreases monotonically with the increase of the strain (see Fig. 4).

The increase of the band valley degeneracy may be beneficial to the thermoelectric performance. In the following, the electronic transport coefficients of the HfS$_{2}$ monolayer under different strains are calculated by using the semiclassical Boltzmann theory. Since the system becomes unstable when the strain reaches 10\%, we only consider the condition when the strain is not larger than 9\%. Figure 5(a) shows the calculated absolute values of the Seebeck coefficients at 300 K as a function of the doping concentration. We can see that for the p-type doping, the Seebeck coefficient first increases as the increase of the strain, reaches its maximum at the strain of 6\%, and then slightly decreases when further increasing the strain. For the n-type doping, however, the absolute value of the Seebeck coefficient increases monotonically as a function of the strain, thus it reaches the maximum at the strain of 9\%. If we notice the energy difference between the band valleys (see Fig. 4), we can find that the Seebeck coefficient increases as the absolute value of the valley energy difference is decreased. When $\Delta_1$  decreases to zero, the degeneracy of the valence band valleys becomes the largest, and thus the absolute value of the Seebeck coefficient of the p-type doping reaches the maximum.

The trend of the electrical conductivity as a function of the strain (see Fig. 5(b)) is just opposite to that of the Seebeck coefficient. Here the relaxation time $\tau$ is inserted as a parameter. The electrical conductivity $\sigma/\tau$ is generally decreased by the applied strain, which is detrimental to the PF. Whether the PF will be improved by the strain or not is determined by the balance between the Seebeck coefficient and electrical conductivity. In Fig. 5(d), we plot the PF (with $\tau$ inserted) at 300 K as a function of the doping concentration under different strains. For the p-type doping, the peak value of the PF first increases as increasing the strain, reaches the maximum at the strain of 6\%, and then decreases slightly when the strain is further increased. For the n-type doping, however, the PF increases monotonically with the increase of the strain. The tread of the PF as a function of the strain is the same as that of the Seebeck coefficient, indicating that the negative effect of the strain on the electrical conductivity is overweighed by the increase of the Seebeck coefficient. Therefore, the PF of the HfS$_{2}$ monolayer can be greatly improved by the valley engineering through the method of strain.

\begin{table}[h]
\small
\caption{\label{Table1-relaxation time}Effective mass ($m^*$), carrier mobility ($\mu$), relaxation time ($\tau$) at 300 K in the zigzag and armchair directions of the unstrained and 6\% strained HfS$_{2}$ monolayers.}

\begin{tabular}{ccccccc}
\hline
~&~&~& $m^*$ & $\mu$ & $\tau$ \tabularnewline
~&~&~& ($m_e$) & (cm$^2$V$^{-1}$s$^{-1}$) & ($10^{-13}$ s) \tabularnewline
\hline
Unstrained & Zigzag & $h$ & $-$0.26 & 1141.6 & 1.68 \tabularnewline
~&~&  $e$ & 0.23 & 4774.3 & 6.35  \tabularnewline
~& Armchair & $h$ & $-$0.25 & 1219.6 & 1.76 \tabularnewline
~&~& $e$ & 2.26 & 502.9 &6.45 \tabularnewline
Strained & Zigzag & $h$ & $-$0.28 & 1801.1 & 2.88 \tabularnewline
~&~&  $e$ & 0.31 & 1021.9 & 1.81 \tabularnewline
~& Armchair & $h$ & $-$0.28 & 1756.8 & 2.77 \tabularnewline
~&~& $e$ & 4.21 & 94.4 & 2.26 \tabularnewline
\hline
\end{tabular}
\end{table}

As for the electronic thermal conductivity, we can see from Fig. 5(c) that the topology of $\kappa_e/\tau$ as a function of the carrier concentration is the same as that of the electrical conductivity, since it is calculated based on $\kappa_e =L\sigma T$. The electronic thermal conductivity is reduced by the strain, which is another beneficial factor to the thermoelectric performance.

As mentioned above, within our method, the electrical conductivity and therefore the PF can only be calculated with the relaxation time $\tau$ inserted as a parameter. The relaxation time is determined by $\mu=e\tau/m^*$, where $\mu$ and $m^*$ are the carrier mobility and effective mass, respectively. Details of calculating the carrier mobility can be found in the ESI.$\dag$ Since at the strain of 6\%, the degeneracy of the valence band valleys reaches the maximum, in the following, we only focus on this strain condition. The calculated $m^*$ and room-temperature $\mu$ and $\tau$ of the unstrained and 6\% strained HfS$_{2}$ monolayer are summarized in Table 1. For the unstrained HfS$_{2}$ monolayer, the effective mass $m^*$ as well as the mobility of the hole are highly isotropic due to the isotropic band dispersion near the VBM. However, for the electron, the effective mass along the zigzag direction is much smaller than that along the armchair direction, since near the CBM, the band dispersion along the $M$-$K$ direction (zigzag direction in real space) is much steeper than that along the $M$-$\Gamma$ direction (armchair direction in real space). The calculated electron mobility along the zigzag direction is as high as 4774.3 cm$^2$V$^{-1}$s$^{-1}$, which is much larger than that of the MoS$_{2}$ monolayer.\cite{MoS2-mobility} The relaxation time of the hole is much smaller than that of the electron. However, for both hole and electron, the difference of the relaxation time $\tau$ between the zigzag and armchair directions is very small, so we will use the averaged $\tau$ along the two directions to evaluate the thermoelectric performance.  When the strain of 6\% is applied, for both the zigzag and armchair directions, the effective mass $m^*$ of the hole are nearly unchanged, while the carrier mobility $\mu$ and relaxation time $\tau$ are slightly increased. The effective mass of electron is however increased, both along the zigzag and armchair directions, because the band dispersion near the CBM becomes flatter when the strain is applied (see Fig. 3). The carrier mobility and the relaxation time of electron are significantly decreased compared with those of the unstrained HfS$_{2}$ monolayer.

\begin{figure}[ht]
\centering
\includegraphics[width=0.9\columnwidth]{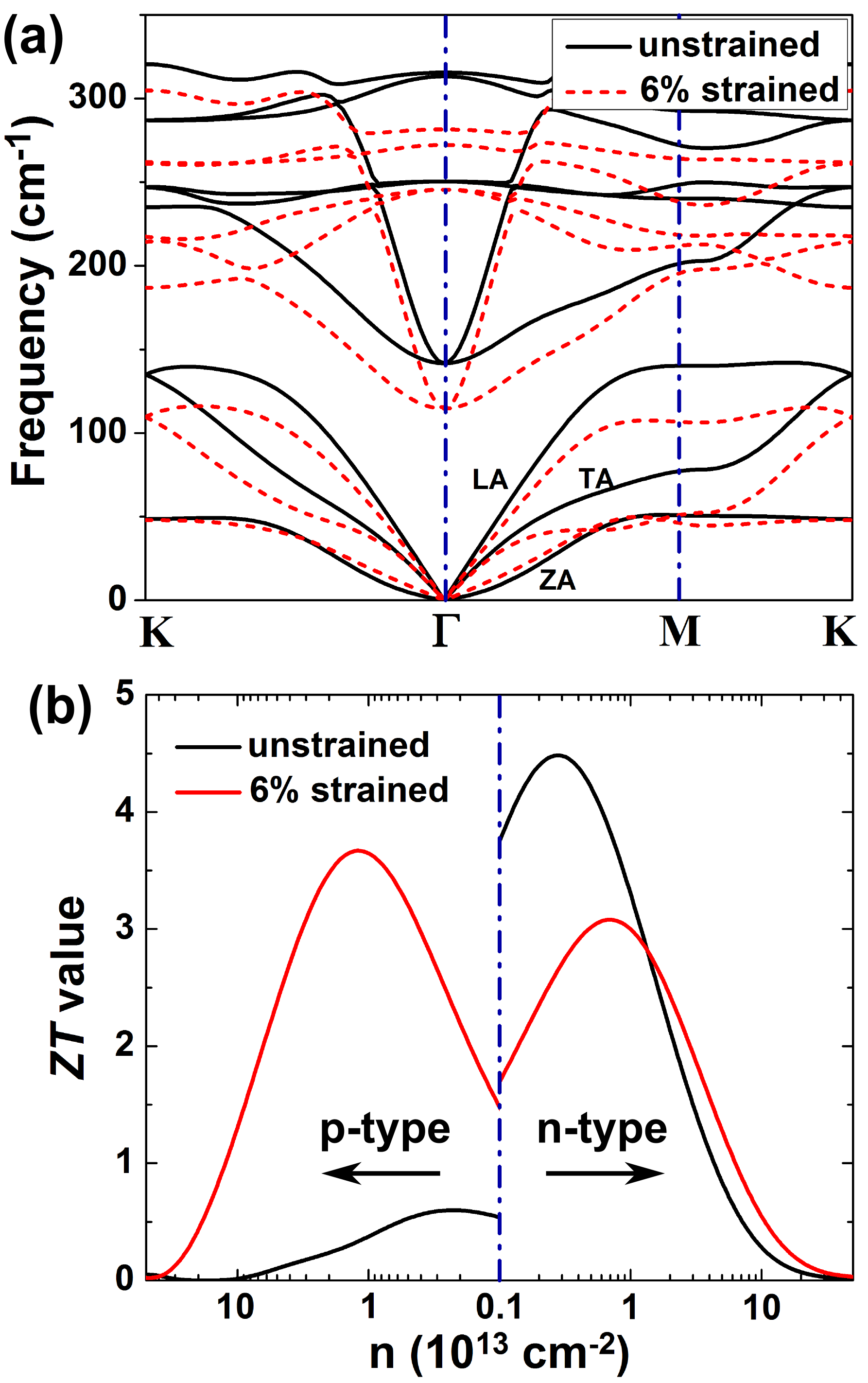}\caption{\label{fig6-ZT value}(a) Phonon spectra and (b) ZT value at 300 K as a function of the carrier concentration for the unstrained and 6\% strained HfS$_{2}$ monolayers. The left and right panels of (b) represent the p- and n-type doping, respectively.}
\end{figure}

\begin{table*}
\centering
\small
\caption{\label{Table1-relaxation time}Optimal doping concentration ($n$) and the corresponding Seebeck coefficient ($S$), electrical conductivity ($\sigma$), electronic and lattice thermal conductivity ($\kappa_e$ and $\kappa_l$), and $ZT$ value of the unstrained and 6\% strained HfS$_{2}$ monolayers at 300 K for the p- and n-type doping.}

\begin{tabular}{ccccccccccccccc}
\hline
~&~&~&~& $n$ &~& $S$ &~& $\sigma$ &~& $\kappa_e$ &~& $\kappa_l$ &~& $ZT$ \tabularnewline
~&~&~&~& (10$^{12}$ cm$^{-2}$) &~& ($\mu$V/K) &~& (10$^5$ $\Omega$$^{-1}$m$^{-1}$) &~& (Wm$^{-1}$K$^{-1}$) &~& (Wm$^{-1}$K$^{-1}$) ~&~& \tabularnewline
\hline
Unstrained &~& p-type &~& 2.25 &~& 197.8 &~& 1.94 &~& 0.87 &~& 2.92 &~& 0.60 \tabularnewline
~&~& n-type &~& 2.75 &~& 360.5 &~& 6.98 &~& 3.14 &~& 2.92 &~& 4.48 \tabularnewline
Strained &~& p-type &~& 12.1 &~& 336.6 &~& 4.67 &~& 2.10 &~& 2.22 &~& 3.67 \tabularnewline
~&~& n-type &~& 7.07 &~& 316.1 &~& 4.24 &~& 1.91 &~& 2.22 &~& 3.08 \tabularnewline
\hline
\end{tabular}
\end{table*}

To evaluate the figure of merit, we have calculated the lattice thermal conductivity $\kappa_p$, which is 2.92 Wm$^{-1}$K$^{-1}$ at 300 K for the unstrained HfS$_{2}$ monolayer. The HfS$_{2}$ monolayer has much smaller lattice thermal conductivity than those of the Mo/W based TMDC monolayers,\cite{MoS2-monolayer-kappa,WS2-monolayer-kappa} the same as the case of the bulk systems.\cite{HfS2-bulk,MoS2-bulk-kappa,WS2-bulk-kappa} Moreover, when under the external strain of 6\%, the room-temperature $\kappa_p$ is further suppressed to be 2.22 Wm$^{-1}$K$^{-1}$, due to the phonon softening of transverse and longitudinal acoustic (TA and LA) modes, as well as the reduced gap between the acoustical and optical phonon bands (see Fig. 6(a)). The very small lattice thermal conductivity of the HfS$_{2}$ monolayer indicating that this system may have much improved thermoelectric performance.

Combining all the calculated coefficients together, we plot in Fig. 6(b) the room temperature $ZT$ value as a function of the carrier concentration. When no strain is applied, a $ZT$ value as high as 4.48 is achieved for the n-type doped system, which is significantly improved compared with the corresponding bulk.\cite{HfS2-bulk} The $ZT$ value of the n-type doped system is much larger than that of the p-type doped one. The HfS$_{2}$ monolayer can be used as an excellent n-type thermoelectric material. However, in the fabrication of thermoelectric modules, both the p- and n-legs are needed. Interestingly, we find that the $ZT$ value of the p-type doped system is significantly increased by the applied strain, from 0.60 for the unstrained HfS$_{2}$ monolayer to 3.67 for the 6\% strained one. However, for the n-type doping, in spite of the increased peak value of the power factor (with relaxation time inserted, see Fig. 5(d)), the $ZT$ value is decreased by the external strain, from 4.48 for the unstrained system to 3.08 for the strained one. This is because that when inserting the thermal conductivity to evaluate the $ZT$ value, the optimal doping concentration gets decreased, moving to the region where the PF of the strained system is slightly smaller than that of the unstrained one. Although slightly decreased, the room-temperature $ZT$ value is still above 3.0, reaching the requirement of the thermoelectric application.\cite{ZT-require}

In Table 2, we summarize the maximum $ZT$ values of the unstrained and 6\% strained HfS$_{2}$ monolayer, with the optimal doping concentration and the corresponding Seebeck coefficient, electrical conductivity, electronic and lattice thermal conductivities included. For the p-type doping, we can see that at the optimal doping concentration, where the maximum $ZT$ value is obtained, both the Seebeck coefficient and the electrical conductivity is increased by the strain, while the lattice thermal conductivity is decreased, which lead to the significantly increased $ZT$ value. For the n-type doping, at the optimal doping concentration, although the electronic and lattice thermal conductivities are decreased, the Seebeck coefficient and the electrical conductivity are also decreased, resulting in the decreased $ZT$ value for the 6\% strained system.

\section{Conclusion}

In conclusion, we have investigated the electronic, phonon, and thermoelectric properties of the HfS$_{2}$ monolayer. The band valleys of this 2D material can be effectively engineered by the external strain. At the strain of 6\%, the three valence band valleys converge in energy and the degeneracy of the valleys reaches the maximum, while the energy difference of the conduction band valleys decreases monotonically as the strain increases up to 9\%, the largest strain the HfS$_{2}$ monolayer can withstand. Although the electrical conductivity is decreased by the strain, the increase of the Seebeck coefficient overweighs the decrease of the electrical conductivity and thus the peak value of the power factor (with relaxation time inserted) increases monotonically with the decrease of the energy difference among the band valleys. At the strain condition of 6\%, the maximum room-temperature $ZT$ value of 3.67 can be achieved for the p-type doped system, which is five times larger than that of the unstrained one. Our results indicate that the thermoelectric performance of the HfS$_{2}$ monolayer can be greatly improved by the band valley engineering through the method of strain.

In experiments, we can introduce strain to a 2D material in various ways. Traditionally, strain can arise from the lattice mismatch between epitaxial thin films and substrates.\cite{mismatch} In recent years, with the progress in nanotechnology, one could transfer a 2D film to a soft supporting substrate and apply strain to the film by either stretching\cite{stretching1,stretching2} or bending\cite{bending1,bending2} the substrate. By using this method, extremely large strain could be achieved. For example, strain as high as 30\% has been realized in graphene.\cite{stretching1,bending2} These methods can be readily transferred to the HfS$_{2}$ monolayer, which deserves further study in the future experiments.

\section{Acknowledgement}

This work was supported by the National Key Research and Development Program under Contract No. 2016YFA0300404, National Natural Science Foundation of China under Contracts No. 11404340, 11274311, U1232139, 11674326 and 11574108, the Anhui Provincial Natural Science Foundation under Contract No. 1408085MA11, the China Postdoctoral Science Foundations (Grant No. 2014M550352 and 2015T80670). The calculation was partially performed at the Center for Computational Science, CASHIPS.


\begin{thebibliography}{10}

\bibitem{MS1} L. D. Hicks and M. S. Dresselhaus, {\it{Phys. Rev. B}}, 1993, \textbf{47}, 12727$-$12731.

\bibitem{MS2} L. D. Hicks and M. S. Dresselhaus, {\it{Phys. Rev. B}}, 1993, \textbf{47}, 16631$-$16634.

\bibitem{synthesis1} J. N. Coleman, M. Lotya, A. O'Neill, S. D. Bergin, P. J. King, U. Khan, K. Young, A. Gaucher, S. De, R. J. Smith, I. V. Shvets, S. K. Arora, G. Stanton, H.-Y. Kim, K. Lee, G. T. Kim, G. S. Duesberg, T. Hallam, J. J. Boland, J. J. Wang, J. F. Donegan, J. C. Grunlan, G. Moriarty, A. Shmeliov, R. J. Nicholls, J. M. Perkins, E. M. Grieveson, K. Theuwissen, D. W. McComb, P. D. Nellist and V. Nicolosi, {\it{Science}}, 2011, \textbf{331}, 568$-$571.

\bibitem{synthesis2} R. J. Smith, P. J. King, M. Lotya, C. Wirtz, U. Khan, S. De, A. O'Neill, G. S. Duesberg, J. C. Grunlan, G. Moriarty, J. Chen, J. Wang, A. I. Minett, V. Nicolosi and J. N. Coleman, {\it{Adv. Mater.}}, 2011, \textbf{23}, 3944$-$3948.

\bibitem{valley1} G. D. Mahan, {\it{Solid State Physics}} (eds H. Ehrenreich and F. Spaepen), Vol. 51, 81-157 (Academic, 1998).

\bibitem{valley2} H. J. Goldsmid, {\it{Thermoelectric Refrigeration}} (Plenum, 1964).

\bibitem{valley3} O. Rabin, Y.-M. Lin and M. S. Dresselhaus, {\it{Appl. Phys. Lett.}}, 2001, \textbf{79}, 81$-$83.

\bibitem{valley-bulk1} Y. Pei, X. Shi, A. LaLonde, H. Wang, L. Chen and G. J. Snyder, {\it{Natuer}}, 2011, \textbf{473}, 66$-$69.

\bibitem{valley-bulk2} W. Liu, X. Tan, K. Yin, H. Liu, X. Tang, J. Shi, Q. Zhang and C. Uher, {\it{Phys. Phys. Lett.}}, 2012, \textbf{108}, 166601.

\bibitem{valley-bulk3} X. J. Tan, W. Liu, H. J. Liu, J. Shi, X. F. Tang and C. Uher, {\it{Phys. Rev. B}}, 2012, \textbf{85}, 205212.

\bibitem{TMDC1} M. N. Ali, J. Xiong, S. Flynn, J. Tao, Q. D. Gibson, L. M. Schoop, T. Liang, N. Haldolaarachchige, M. Hirschberger, N. P. Ong and R. J. Cava, {\it{Nature}}, 2014, \textbf{514}, 205$-$208.

\bibitem{TMDC2} B. Sipos, A. F. Kusmartseva, A. Akrap, H. Berger, L. Forr\'{o} and E. Tuti\v{s}, {\it{Nat. Mater.}}, 2008, \textbf{7}, 960$-$965.

\bibitem{TMDC3} A. H. Castro Neto, {\it{Phys. Rev. Lett.}}, 2001, \textbf{86}, 4382$-$4385.

\bibitem{TiS2-bulk} H. Imai, Y. Shimakawa and Y. Kubo, {\it{Phys. Rev. B}}, 2001, \textbf{64}, 241104(R).

\bibitem{TiS2-layer} R.-Z. Zhang, C.-L. Wan, Y.-F. Wang and K. Koumoto, {\it{Phys. Chem. Chem. Phys.}}, 2012, \textbf{14}, 15641$-$15644.

\bibitem{HfS2-bulk} G. Yumnam, T. Pandey and A. K. Singh, {\it{J. Chem. Phys.}}, 2015, \textbf{143}, 234704.

\bibitem{MoS2-bulk-kappa} A. N. Gandi and U. Schwingenschl\"{o}gl, {\it{Europhys. Lett.}}, 2016, \textbf{113}, 36002.

\bibitem{WS2-bulk-kappa} A. N. Gandi and U. Schwingenschl\"{o}gl, {\it{Chem. Mater.}}, 2014, \textbf{26}, 6628$-$6637.

\bibitem{ZT-require} B. C. Sales, {\it{Science}}, 2002, \textbf{295}, 1248$-$1249.

\bibitem{ABINIT1} X. Gonze, J.-M. Beuken, R. Caracas, F. Detraux, M. Fuchs, G.-M. Rignanese, L. Sindic, M. Verstraete, G. Zerah, F. Jollet, M. Torrent, A. Roy, M. Mikami, P. Ghosez, J.-Y. Raty and D. C. Allan, {\it{Comput. Mater. Sci.}}, 2002, \textbf{25}, 478$-$492.

\bibitem{ABINIT2} X. Gonze, G.-M. Rignanese, M. Verstraete, J.-M. Beuken, Y. Pouillon, R. Caracas, F. Jollet, M. Torrent, G. Zerah, M. Mikami, P. Ghosez, M. Veithen, J.-Y. Raty, V. Olevano, F. Bruneval, L. Reining, R. Godby, G. Onida, D. R. Hamann and D. C. Allan, {\it{Z. Kristallogr.}}, 2005, \textbf{220}, 558$-$562.

\bibitem{ABINIT3} X. Gonze $et$ $al.$, {\it{Comput. Phys. Commun.}}, 2009, \textbf{180}, 2582$-$2615.

\bibitem{PBE} J. P. Perdew, K. Burke and M. Ernzerhof, {\it{Phys. Rev. Lett.}}, 1996, \textbf{77}, 3865$-$3868.

\bibitem{Boltzmann} G. K. H. Madsen and D. J. Singh, {\it{Comput. Phys. Commun.}}, 2006, \textbf{175}, 67$-$71.

\bibitem{rigid-band} E. A. Stern, {\it{Phys. Rev.}}, 1967, \textbf{157}, 544$-$551.

\bibitem{Lorenz} R. Venkatasubramanian, E. Siivola, T. Colpitts and B. O'Quinn, {\it{Nature}}, 2001, \textbf{413}, 597$-$602.

\bibitem{vasp1} G. Kresse and J. Hafner, {\it{Phys. Rev. B}}, 1993, \textbf{47}, 558$-$561.

\bibitem{vasp2} G. Kresse and J. Hafner, {\it{Phys. Rev. B}}, 1994, \textbf{49}, 14251$-$14269.

\bibitem{vasp3} G. Kresse and J. Furthm\"{u}ller, {\it{Comput. Mater. Sci.}}, 1996, \textbf{6}, 15$-$50.

\bibitem{phonopy} A. Togo and I. Tanaka, {\it{Scr. Mater.}}, 2015, \textbf{108}, 1$-$5.

\bibitem{ShengBTE} W. Li, J. Carrete, N. A. Katcho and N. Mingo, {\it{Comput. Phys. Commun.}}, 2014, \textbf{185}, 1747$-$1758.

\bibitem{bulk-structure} D. L. Greenaway and R. Nitsche, {\it{J. Phys. Chem. Solids}}, 1965, \textbf{26}, 1445$-$1458.

\bibitem{few-layer HfS2} T. Kanazawa, T. Amemiya, A. Ishikawa, V. Upadhyaya, K. Tsuruta, T. Tanaka, and Y. Miyamoto, {\it{Sci. Rep.}}, 2016, \textbf{6}, 22277.

\bibitem{monolayer-phonon} J. Kang, H. Sahin, and F. M. Peeters, {\it{Phys. Chem. Chem. Phys.}}, 2015, \textbf{17}, 27742$-$27749.

\bibitem{MoS2-mobility} B. Radisavljevic and A. Kis, {\it{Nat. Mater.}}, 2013, \textbf{12}, 815$-$820.

\bibitem{MoS2-monolayer-kappa} Y. Cai, J. Lan, G. Zhang, and Y.-W. Zhang, {\it{Phys. Rev. B}}, 2014, \textbf{89}, 035438.

\bibitem{WS2-monolayer-kappa} D. Wickramaratne, F. Zahid, and R. K. Lake, {\it{J. Chem. Phys.}}, 2014, \textbf{140}, 124710.

\bibitem{mismatch} G. Gao, S. Jin and W. Wu, {\it{Appl. Phys. Lett.}}, 2007, \textbf{90}, 012509.

\bibitem{stretching1} K. S. Kim, Y. Zhao, H. Jang, S. Y. Lee, J. M. Kim, K. S. Kim,
J.-H. Ahn, P. Kim, J.-Y. Choi and B. H. Hon, {\it{Nature}}, 2009, \textbf{457}, 706$-$710.

\bibitem{stretching2} M. A. Bissett, S. Konabe, S. Okada, M. Tsuji, and H. Ago, {\it{ACS Nano}}, 2013, \textbf{7}, 10335$-$10343.

\bibitem{bending1} S.-I. Park, J.-H. Ahn, X. Feng, S. Wang, Y. Huang and J. A. Rogers, {\it{Adv. Funct. Mater.}}, 2008, \textbf{18}, 2673$-$2684.

\bibitem{bending2} Y. Wang, R. Yang, Z. Shi, L. Zhang, D. Shi, E. Wang, and G. Zhang, {\it{ACS Nano}}, 2011, \textbf{5}, 3645$-$3650.

\end{thebibliography}
\end{document}